\documentclass[a4paper,fleqn,usenatbib]{mn2e}
\usepackage{graphicx}
\usepackage{amsmath}
\usepackage{amssymb}
\usepackage{upgreek}
\usepackage[multidot]{grffile}
\usepackage{hyperref}

\title[The Period-Width relationship for radio pulsars]
{The Period-Width relationship for radio pulsars revisited}

\author[Johnston \& Karastergiou]  {Simon Johnston$^{1}$\thanks{email: Simon.Johnston@csiro.au} and A. Karastergiou$^{2,3,4}$
\\
$^{1}$CSIRO Astronomy and Space Science, Australia Telescope National Facility, PO Box 76, Epping, NSW 1710, Australia\\
$^{2}$Oxford Astrophysics, Denys Wilkinson Building, Keble Road, Oxford, OX1 3RH, UK.\\
$^{3}$Physics Department, University of the Western Cape, Cape Town 7535, South Africa\\ 
$^{4}$Department of Physics and Electronics, Rhodes University, PO Box 94, Grahamstown 6140, South Africa
}
\date{Accepted \today. Received \today; in original form \today}

\pubyear{2018}

\begin{document}
\label{firstpage}
\pagerange{\pageref{firstpage}--\pageref{lastpage}} 
\maketitle

\begin{abstract}
In the standard picture of radio pulsars, the radio emission arises
from a set of open magnetic field lines, the extent of which is primarily
determined by the pulsar's spin period, $P$, and the emission height.
We have used a database of parameters from 600 pulsars to show that
the observed profile width, $W$, follows $W\propto P^{-0.3}$
albeit with a large scatter, emission occurs
from heights below 400~km and that the beam is underfilled.  Furthermore,
the prevalence in the data for long period pulsars to have relatively wide
profiles can only be explained if the angle between the magnetic
and rotation axis decays with time.
\end{abstract}


\section{Introduction}
Soon after the discovery of radio pulsars it was realised that they
were rapidly rotating, highly magnetised neutron stars. The high
rotation rate means that not all the magnetic field lines are closed
and radiation streams out from the open field lines resulting in a
regular pulse of radio emission once per rotation as seen from
Earth. The extent of the open field lines depends primarily on the
rotation rate and hence there is an expectation of a relationship
between the pulsar spin period, $P$, and the opening angle of the cone
of the bundle of open field lines. However, the
observed pulse width, $W$, depends on a plethora of factors including the
height of the radio emission above the surface of the star, the
geometry of the star, the viewing angle and the extent to which the
beam is filled with emitting regions.

Much can therefore be learned simply by measuring the width of the
pulse profile and indeed there is a long tradition of such work in the
literature with key papers in the 1990s
\citep{ran90,ran93,gks93,kwj+94,kxl+98,gl98,tm98} and again over the
last decade \citep{mr02,wj08b,ycbb10,mg11,mgm12,sbm+18}. 
In this paper we revisit
this question using a set of 600 pulsars observed with the Parkes
telescope \citep{jk18} and a novel technique for measuring pulse widths 
down to 10\% of the peak level. Section~2 provides a review of the factors 
which go into the observed pulse width.
Section~\ref{dataset} introduces the dataset
and the analysis technique, the results are given in
Section~\ref{results}.  Finally Section~5 discusses the implications
of the results in the context of a simulation of the population.

\section{What factors determine the width of a radio pulsar profile?}
The simplest assumptions to make are that the magnetic field structure
is dipolar, that the radio emission occurs at a height $h_{em}$ and
that emission entirely fills the (circular) open field line region.
In this case, the half-opening angle, $\rho$, of the cone of radio
emission is given by
\begin{equation}
\rho = s\,\,\, \sqrt{\frac{9 \,\, \pi \,\,\, h_{\rm em}}{2\,\,\, P\,\,\, c}}
\label{rho}
\end{equation}
with $P$ the pulsar spin period and $c$ the speed of light \citep{ran90}.
The parameter $s$ is the ratio of the emission longitude to the size
of the polar cap. For simplicity, it is generally taken to be 1.0.

Using these simple ideas, the width of the pulse profile that one measures, 
$W$, is then simply given by a combination of $\rho$ and the 
geometry as follows:
\begin{equation}
{\rm cos}\rho = {\rm cos}\alpha\,\, {\rm cos}\zeta\,\, +\,\, {\rm sin}\alpha\,\, {\rm sin}\zeta\,\, {\rm cos}(W/2)
\label{W}
\end{equation}
where $\alpha$ is the inclination angle between the rotation and
magnetic axis and $\zeta = \alpha + \beta$ where $\beta$ is the angle
between the magnetic axis and the observers line of sight
\citep{ggr84}.  This equation shows that for $\alpha=90\degr$ and
$\beta=0\degr$, then $W=2\rho$.  Measured widths can be narrower than
$2\rho$ for high values of $\alpha$ when the line of sight cuts near
the bottom or top of the cone (i.e. high $|\beta|$).  Conversely, measured
widths can be significantly larger than $2\rho$ for low values of
$\alpha$ where the line-of-sight can remain within the emission cone
for a significant fraction of the spin period.

In the following subsections we examine some of our assumptions in
more detail.

\subsection{Emission Height}
It can be seen from Equation~\ref{rho} that if $h_{em}$ is a fixed
value for all pulsars then $\rho \propto P^{-1/2}$, but if $h_{em}$ is
a fixed fraction of the light cylinder radius than $\rho$ will be
independent of $P$. Indeed, the literature to date prefers the former,
with \cite{kwj+94,kxl+98,ran93} showing $\rho \propto P^{-1/2}$ and
others \citep{gl98,wj08b} showing a somewhat flatter relationship.

More complicated arrangements are also possible. \citet{gg03} showed
that emission heights are large on the outside of the polar cap but
smaller towards the middle of the polar cap.  \citet{kj07} showed that
emission heights for young pulsars can be large but restricted to a
narrow range, whereas older pulsars can have a much wider range of
possible emission heights.

Emission heights can, in principle, also be computed using a method first 
developed by \citet{bcw91}. This relies on the determination of the offset 
between the centroid of the pulse profile and the location of the steepest 
gradient of the swing of the position angle of the linear polarization.
Unfortunately, as \citet{wj08b} showed, the agreement between heights
determined using this method and those computed from the pulse profile
width is rather poor.

We note that some authors invoke the \citet{rs75} model to explain the
significance of the $\rho \propto P^{-1/2}$ observational
result. However, there is as yet no theory which links the emission
height to the pulsar parameters and so there is no {\it a priori}
reason to expect that $\rho \propto P^{-1/2}$.

\subsection{Emission from outside the polar cap}
In the standard picture of radio emission, emission from outside the
conventionally-defined open field lines is not expected.
Some recent evidence, however, seems
to indicate that this might be violated for the short-period, highly
energetic pulsars. For the interpulse pulsar PSR~J1057--5526, \cite{ww09}
showed that emission from the main pulse must arise outside the open
field lines. \cite{rwj15a,rwj15b} showed the same puzzling results for a number
of $\gamma$-ray loud pulsars; in order to reconcile the $\gamma$-ray
emission with the wide radio profiles, emission from outside the open
field lines must be present. They showed that there is a linear dependence
between $s$ and $\alpha$ with pulsars with lower $\alpha$ having larger
values of $s$, and indeed for $\alpha<30\degr$ then $s>2$.

\subsection{Beam structure}
We know that the open field lines are not uniformly filled with radio
emission, because pulsars show a wide and bewildering variety of
structures in their pulse profiles. In the empirical models of Rankin
and collaborators, the beams have either one or two `conal' rings of
emission in addition to emission from near the centre (`core') of the
beam (see e.g. Figure~1 in \citealt{ran93}). Results of model fitting
imply that the emission comes from a low height, independent of pulsar
period \citep{mr02} and the cones only cover part of the available
beam. Outer cones appear to have a higher emission height than inner
cones \citep{mr02,gg03}.  Some authors, most recently \citet{mg11}, 
\citet{mgm12} and \citet{sbm+18}, have attempted to fit the period-width
relationship to core and cone components separately and we will return 
to this issue later.

This organised picture finds its contrast in the work of \cite{lm88}
who present evidence for the patchy nature of the pulsar beam rather
than for organised conal structures. In their model, `patches' of
emission are spread at random over the polar cap and this can lead to
the so-called `partial cone' pulsars where only one edge of the beam
is apparent.  According to \cite{lm88}, therefore, the observed widths
may not be a reflection of the entire polar cap due to the patchiness
(although see \citealt{mr11} for a rebuttal).

For the pulsars with the highest values of $\dot{E}$, \cite{jw06}
noticed that many of the profiles are `wide doubles' with overall
widths of 100$\degr$ or more and with the trailing component
dominating. Such pulsars stick out in the $\dot{E} - W$ plane as noted 
by \cite{wj08b}. This may in turn imply that the beam structure of the 
energetic pulsars is different from that of older pulsars \citep{rmh10}.

We use $f$ to parameterise the filling fraction of the beam along the line
of sight cut through the polar cap. $f=1$ implies that emission occurs
(at least) on the leading and trailing edges of the polar cap so that the
measured width would be the same as that expected from Equation~\ref{W}.
For, say $f=0.1$, the beam is patchy with only 10\% illuminated so that
the measured width is significantly smaller than expected.

\subsection{Beam circularity}
Much of the above discussion is predicated on beams that are circular
and bounded by the open field lines. This is however an assumption
built into the models. Over the years, there has been discussion of
either longitudonal \citep{mck93,gan04} or latitudonal \citep{nv83}
compression of the beams. These are relatively minor perturbations of
the circular assumption.  The possibility of elliptical beams has
been mooted to explain the sub-pulse bi-drifting \citep{svl17,ww17}
but the applicability to the population as a whole is not clear.
At the same time, observations of precession
of pulsars have allowed us a mapping of the radio beam the results of
which are intriguing.  PSRs J1141--6545 \citep{mks+10} and J1906+0746
\citep{dkc+13} have beams which are very underfull in longitude but
filled in latitude and similar though less extreme results are also
seen in PSRs B1913+16 \citep{cw08} and B1534+12 \citep{fst14}. As a
result, \cite{wpz+14} proposed a fan-beam model to produce radially
extended beams. Their model predicts that the width increases
(contrary to the standard circular beam picture) and the intensity
decreases as $\beta$ increases.

\subsection{Distribution of $\alpha$}
The birth distribution of $\alpha$ is also unclear, with many authors
postulating a random value of $\alpha$ as a starting assumption. Note
that a random value of $\alpha$ in the population will not be
reflected in the observed population. This is because orthogonal
rotators have beams which cover a much larger solid angle of sky than
do aligned rotators and so have a greater chance of detection. Hence
the observed population should have a sine-like distribution of $\alpha$.

There is strong evidence that $\alpha$ decays towards zero (alignment of spin and magnetic axes) over the
observable lifetime of a radio pulsar, as presented in \citet{tm98}, \citet{wj08a}, \citet{ycbb10} and \citet[hereafter JK17]{jk17}. This
will affect the observed distribution of pulse widths as a function of
age (or period).  On the other hand, the young pulsar in the Crab
Nebula appears to have $\alpha$ increasing towards orthogonality over
the few decades of observation of the pulsar \citep{lgw+13}, so the
situation for the youngest pulsars is far from clear \citep{rwj15b}.

\subsection{Selection effects}
The observed population of $\sim$2500 pulsars is not representative of
the pulsar population as a whole. In particular, the search techniques
strongly favour pulsars with narrow pulse widths over those with broad
pulses (see e.g. the review in \citealt{vhkr17}).
This is particularly acute at long pulse periods where the red
noise in the Fourier transform of the time series largely precludes finding 
wide, nearly aligned profiles \citep{lbh+15}.
In addition, the fan-beam model of \citet{wpz+14} predicts that the radio 
luminosity is a function of $\beta$, implying
that pulsars discovered at large distances should (preferentially) have
low values of $\beta$ compared to those discovered nearby.

\subsection{Summary}\label{sum2}
Although Equations~\ref{rho} and \ref{W} promise much, we have seen
the dangers in using them blindly. First $h_{em}$ can vary from
$\sim$100~km to several thousand km both for different emission
regions in the one pulsar and from pulsar to pulsar.
Secondly, $s$ appears to be as large as
$\sim$4 in the young pulsars \citep{rwj15b}, but only $\sim$0.7 for
older pulsars \citep{mr02}.
In addition, $h_{em}$ and $s$ can only be disentangled if
one of them can be measured independently through other means.
Thirdly, $\alpha$ and $\zeta$ are difficult to measure for a given
pulsar and the underlying distribution of $\alpha$ and its
time-derivative are unclear. The beam may not be filled and `missing'
parts of the emission profile may lead to an underestimation of the
width of the profile.  Finally, the beam may not be circular and can
either be marginally compressed in longitude or latitude, be elliptical,
or in the shape of a fan-beam.

\begin{figure}
\includegraphics[width=6.5cm,angle=-90]{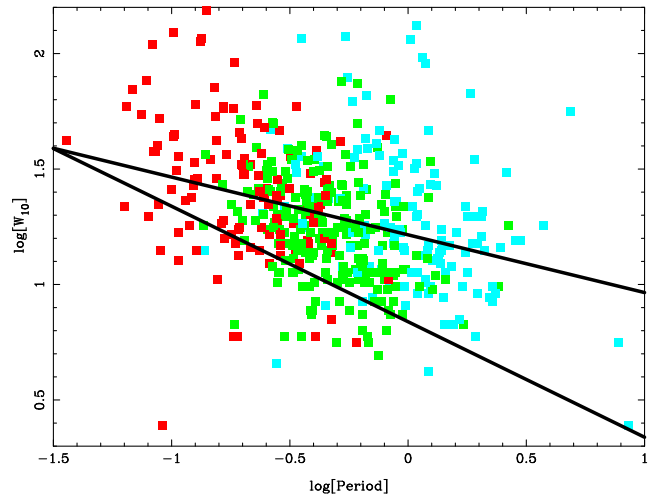}
\caption{Profile widths at 10\% of the peak value versus pulse period
for 475	pulsars at an observing frequency of 1.4~GHz. Red points
are pulsars with $\dot{E} > 10^{34}$~erg\,s$^{-1}$, green points have
$10^{32}$~erg\,s$^{-1} < \dot{E} < 10^{34}$~erg\,s$^{-1}$ and
blue points have $\dot{E} < 10^{32}$~erg\,s$^{-1}$.  The lower-bound
line, $W_{\rm 10}=6.9\degr P^{-0.5}$, used by \citet{gl98} is shown
as is the fit to our data, $W_{\rm 10}=15.8\degr P^{-0.28}$.}
\label{data-all}
\end{figure}

\section{Dataset and Analysis}\label{dataset}
We use the 600 pulsars observed at 1.4~GHz using the Parkes telescope
and described in the paper by \citet{jk18}. This is a homogeneous
sample of southern hemisphere pulsars brighter than $\sim$0.7~mJy. In
that paper, only values of $W_{50}$ (the width of the profile at
50\% of the peak) were listed but we have developed a method for measuring 
$W_{10}$ (the width of the profile at 10\% of the peak) in the presence of 
noise, enabling us to measure accurate values for both high and low peak 
signal-to-noise ratio profiles.

The method relies on the assumption that the data consist of signal
and a white noise term. For simplicity, it is assumed that the
variance of the white noise term is constant across the profile
(homoscedastic), whereas in reality, the variance will be slightly
greater for parts of the profile where the signal is stronger
(heteroscedastic). The signal is assumed to be a smooth function of
pulse phase, and no other assumption is made. We then use the data to
derive a Gaussian Process for each profile, which is a Bayesian,
non-parametric model, in the sense that no assumption is made about
the functional form of the pulse profile with phase
\citep[GP]{roe+12}. The GP requires prior choice of a covariance
function, which governs the way in which the intensity at a given
pulse phase affects and is affected by its surroundings. We use a
squared exponential covariance function, as it satisfies our
smoothness assumption and is infinitely differentiable, providing us
with a means to analytically compute the derivatives of the model:
\begin{equation}\label{cov_function}
k(x_i,x_j) = h^2 \exp
\left[-\left(\frac{x_i-x_j}{\lambda}\right)^2\right],
\end{equation}
where $h$ and $\lambda$ are two hyper-parameters that control the
magnitude and length-scale of the covariance. The full covariance
matrix of the GP is then:
\begin{equation}
\bf{V}(\bf{x},\bf{x}) = \bf{K}(\bf{x},\bf{x}) + \sigma^2\bf{I},
\end{equation}
where $\bf{K}$ is the covariance matrix whose elements are given in
Eq. \ref{cov_function}, and $\bf{V}$ the covariance matrix including
the white noise term mentioned above. In total, this model has three
hyper-parameters, adding $\sigma$ to the aforementioned
two. \citet{roe+12} explain how the hyper-parameters are best
determined, and we follow their guidance in our solution. It is worth
considering for a moment the potential of modeling pulse profiles in
this way. First, there is no requirement for a definition of an
on-pulse and off-pulse region to determine the noise
variance. Secondly, the model can be separated into its two
constituent parts, the signal and the noise, effectively yielding an
optimized version of a noiseless profile, given the data. Thirdly, the
derivatives of the signal model can be computed analytically.
While it is true that the shape of pulsar profiles suggests that a model with
pulse-phase dependent hyper-parameters would be more appropriate than
the model we have used here, in practice, the model performs exceptionally 
well in produce noiseless profiles of very high fidelity.

\begin{figure}
\includegraphics[width=6.5cm,angle=-90]{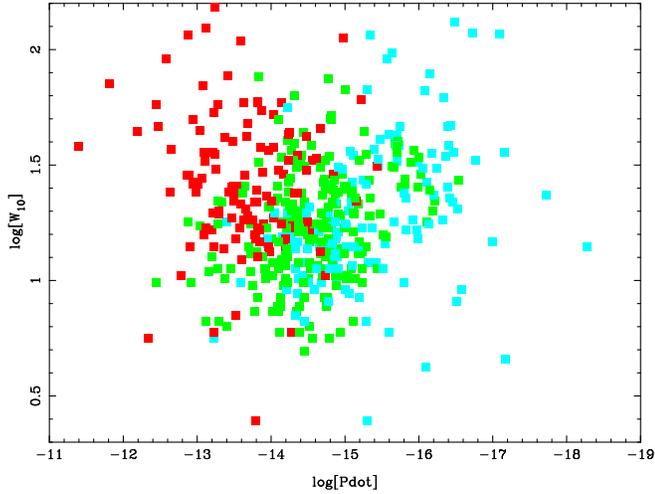}
\caption{Period derivative ($\dot{P}$) versus $W_{10}$ for 475 pulsars.
The colour convention is as per Figure~\ref{data-all}.}
\label{pdot}
\end{figure}
\begin{figure}
\includegraphics[width=6.5cm,angle=-90]{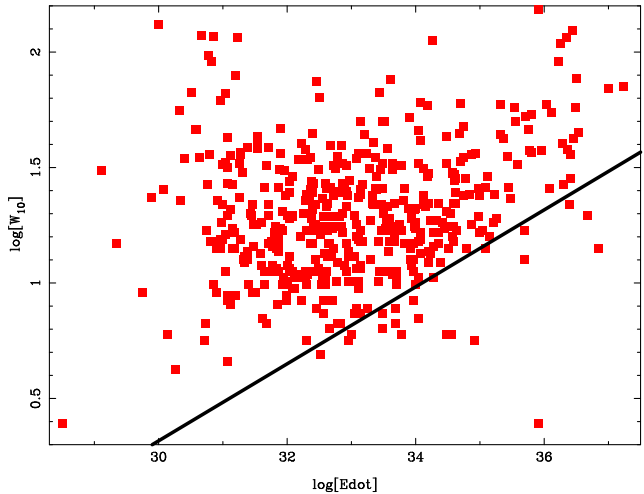}
\caption{Spin-down energy ($\dot{E}$) versus $W_{10}$ for 475 pulsars.
  The straight line has a slope of $-1/6$ as expected if $W \propto
  P^{-0.5}$.}.
\label{edot}
\end{figure}
\begin{figure}
\includegraphics[width=6.5cm,angle=-90]{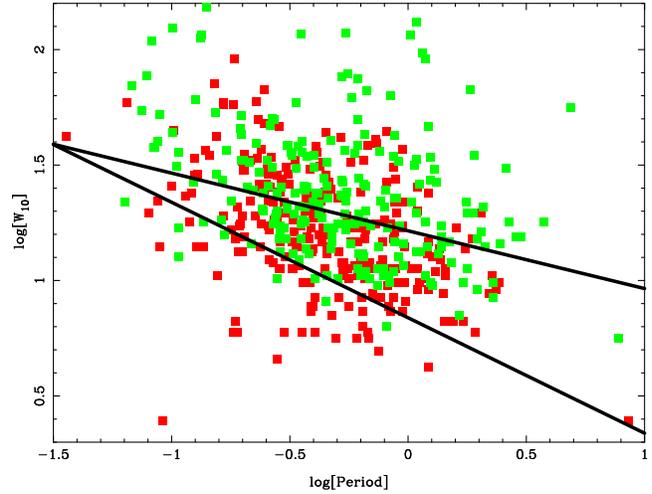}
\caption{Profile widths at 10\% of the peak value versus pulse period
for 254 single pulsars (red) and 232 multiple pulsars (green).}
\label{data-doubles}
\end{figure}
\section{Results}\label{results}
From the 600 pulsars in the sample, we removed 32 which showed
obvious signs of scatter-broadening \citep{jk18} and removed a further 4
pulsars with no measured $\dot{P}$ or $\dot{E}$.
From the remaining 564
pulsars, we measured $W_{10}$ for 475 and were unable to obtain a fit
for 89 pulsars with a measured peak low signal-to-noise ratio smaller than 10.
Finally, for 9 pulsars showing
interpulse emission we separately measured $W_{10}$ for both the main
and interpulses, giving a grand total of 484 measured widths.
Given that all width measurements are conducted on high 
signal-to-noise ratio profiles ($>10$), the error associated with the 
measurement is small compared to the spread in measured widths of the sample.
It is indicative that the width error associated with the 
noisiest profiles corresponds to a few phase bins, or $\sim 1\degr$.

Figure~\ref{data-all} shows $W_{10}$ versus $P$ for these pulsars.
There is a strong anti-correlation between
the parameters and a large scatter about the mean trend. Two lines are
shown in Figure~\ref{data-all}. The first is the lower bound defined
by $W_{10}=6.9\degr P^{-0.5}$ as given in \citet{gl98} and the second is a
straight line fit through the data which yields
$W_{10}=15.8\degr(\pm0.6) P^{-0.28\pm0.03}$. 
The {\bf lower bound} in Figure~\ref{data-all} comes from
the widespread claim in the literature that, for a constant emission
height, the minimum width is given when $\alpha=90\degr$,
$\beta=0\degr$ \citep{ran90,gl98,mgm12}.
However, this is incorrect. First, higher values of
$\beta$ will reduce the width and, secondly, if the polar cap is not
filled and/or the beam is patchy this will also decrease the observed width. 

The pulsar with the narrowest width, PSR~J1028$-$5819 with a spin-period of 
91~ms \citep{kjk+08}, lies more than an order
of magnitude below the \citet{gl98} line. It has a double-peaked
radio profile with a high degree of polarization
and is a $\gamma$-ray emitter; either its radio beam is
extraordinarily underfilled or the line-of-sight just cuts through the
very edge of the beam. The pulsar with the widest (unscattered) profile
is PSR~J1015$-$5719 with a width of 153\degr\ and a spin period of 140~ms.

Figure~\ref{pdot} shows $W_{10}$ versus $\dot{P}$, the derivative of
the spin period.  There is no correlation between the parameters, with
a large scatter in $W_{10}$ at all values of $\dot{P}$.
Figure~\ref{edot} shows $W_{10}$ versus $\dot{E}$, the rotational
spin-down energy of the pulsar. As $\dot{E} \propto P^{-3}\dot{P}$ so if
$W_{10} \propto P^{-1/2}$ and is uncorrelated with $\dot{P}$
then $W_{10} \propto \dot{E}^{1/6}$.
Indeed, at values of $\dot{E} > 10^{33}$~erg\,s$^{-1}$ this slope is
observed in the data albeit with low significance.  Below $\dot{E} \sim
10^{33}$~erg\,s$^{-1}$ however, the measured widths do not continue to
decline and if anything, appear to {\it increase} contrary to
expectations, with a significant number of pulsars with
$\dot{E}<10^{32}$~erg\,s$^{-1}$ and $W_{10}>30\degr$, an order of
magnitude larger than expectations.

\begin{figure}
\includegraphics[width=6.5cm,angle=-90]{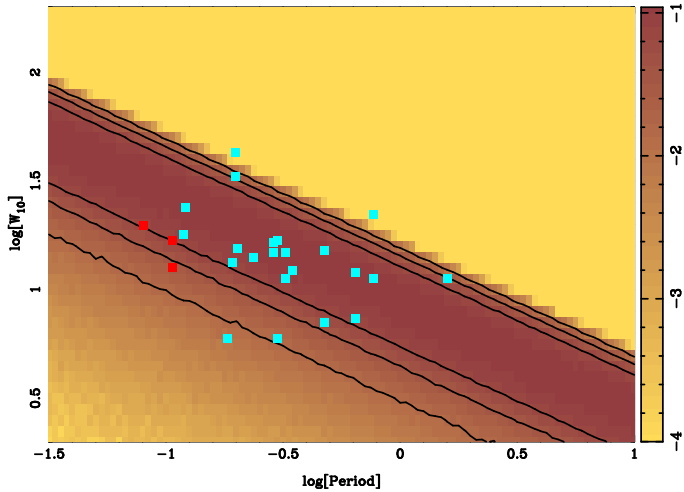}
\caption{Period-width diagram for pulsars showing emission from both
poles. Squares denote the measurements with the colours representing 
$\dot{E}$ as in Figure~\ref{data-all}.  The underlying heat map 
represents the log of the pdf of a simulation which has a fixed 
$\alpha=85\degr$, $h_{em}$ between 200 and 400~km, and the average filling
fraction set to 0.7. Contours denote 0.5, 1.0 and 1.5-$\sigma$ down from the
peak of the heat map.}
\label{ips}
\end{figure}
\subsection{Single versus multiple components}
We sub-divided the 600 pulsars into two main classes. The first class,
the `singles', shows a single, gaussian-like component with no evidence
of multiple components.  The second class, the `multiples', shows
clear evidence for multiple components in the profile.  There are 254
pulsars (52\%) in the first class and 232 (48\%) in the second class with 
the rest (124 pulsars) either too weak or ambiguous to classify.
Figure~\ref{data-doubles} shows
the same plane as Figure~\ref{data-all} for pulsars in these two
classes.  Both classes occupy the entire period space and there is
strong overlap in the widths of the two classes. However, of the 36 pulsars
with $W_{10}>50\degr$, only 8 have a single component.
Conversely, of the 62 pulsars with $W_{10}<10\degr$ only 12 are
multiple component pulsars.
On average,
pulsars with multiple components are 50\% wider than those with
only one component; fitting a straight line through the data yields
$W_{10}=12.3\degr(\pm0.6) P^{-0.34\pm0.04}$,
and for $W_{10}=19.5\degr(\pm0.9) P^{-0.26\pm0.04}$ for
the single and multiple component pulsars respectively.
\begin{figure}
\includegraphics[width=6.5cm,angle=-90]{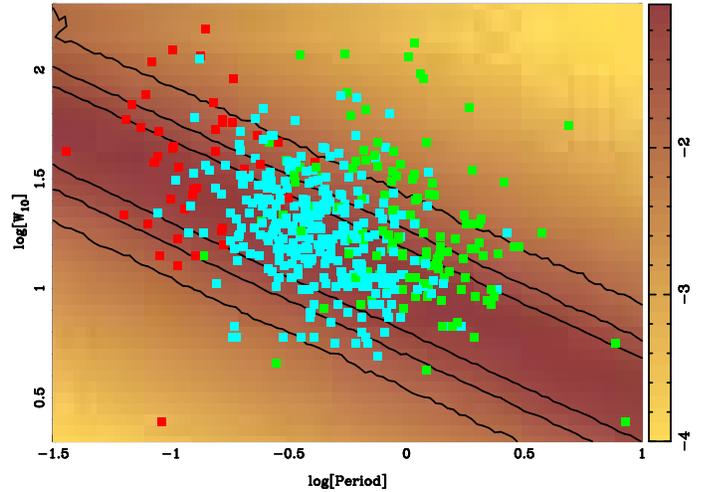}
\caption{Period-width diagram for 475 pulsars. Squares denote the measurements 
with the colours representing $\dot{E}$ as in Figure~\ref{data-all}.
$\alpha$ has a random distribution for all $P$, $h_{em}$ lies between 200 and 
400~km, and the average filling fraction set to 0.7.
Colour coding of the background and the contours are as per Figure~\ref{ips}.}
\label{norook}
\end{figure}

\section{Implications and Discussion}
How then do we overcome the issues outlined in Section~\ref{sum2} in
order to make sense of the observed profile widths presented in
Section~\ref{results}?

\subsection{Simulations}
We create a simulation such that it is possible to obtain a
probability density function (pdf) for the width of a pulsar profile given its
period and a set of free parameters.  The free parameters are:
\begin{itemize}
\item a functional form for the pdf of $h_{em}$ including a possible 
dependence with age or spin-down energy
\item a functional form for the pdf of $s$ including a possible 
dependence of $s$ on $\alpha$
\item a functional form for the pdf of $\alpha$ including a possible 
dependence with age or spin period
\item a functional form for the pdf of $f$, the filling fraction of the beam
\end{itemize}
We retain the simplification of circular beams. This implies that
$\beta$ can be drawn from a flat distribution between $\pm\rho$ once
$\rho$ is determined.  A Monte-Carlo simulation using the free
parameters will then yield a probability density function for $\rho$
via Equation~\ref{rho} and hence $W$ via Equation~\ref{W} for a given
combination of pulsar geometry, period, age and spin-down energy.

\subsection{Interpulses}
The interpulse pulsars are critical to the results for two reasons.
First, $\alpha\approx 90\degr$ thereby simplifying the relationship
between $W$ and $\rho$ (see equation~\ref{W}). Secondly, for these
pulsars we often know the full geometry well (see
e.g. \citealt{kj08,kjwk10}) including $\beta$ and $h_{em}$.  Note that
$\alpha_{IP}=180\degr-\alpha_{MP}$ and
$\beta_{IP}=180\degr-2\alpha_{MP}+\beta_{MP}$ so that in some instances
$\beta_{MP}$ can be very different from $\beta_{IP}$ with the
subsequent implications for a difference in measured width between the
main and interpulse. There is also no a-priori reason why the emission
height of the main and interpulses should be the same.

Generally \citet{kjwk10} find that $h_{em} < 300$~km (except for
PSR~J1057--5552) and that the beams are somewhat underfilled.  In the
dataset used here, there are 19 interpulse pulsars and we can measure
$W_{10}$ for 16 of the main pulses and for 10 of the interpulses. Note
that, unlike in \citet{mgr11} we do not attempt to identify `core'
components in these pulsars but measure the full width ($W_{10}$) of
the profile.  Figure~\ref{ips} shows the period-width plane for these
26 values.  A straight line fit to these data yields
$W_{10} = 10.7\degr(\pm2.1) P^{-0.21\pm0.12}$, thus a similar slope to the 
data as a whole but a narrower width (as expected given that 
$\alpha\approx 90\degr$). If we fix the slope at $-0.5$ then the result is 
$W_{10} = 6.8\degr(\pm0.7) P^{-0.5}$, similar to that found by \citep{gl98}.

We find that the simulation cannot reproduce these widths using only 
a spread in $h_{em}$.  Rather, a combination of filling factor and
heights are required for the simulation to match the data. There
is also additional evidence that the narrow widths seen in e.g.
PSR~J1611--5209 {\bf require} that the beam be underfilled \citep{kjwk10}.
The data can be reproduced with a range in $h_{em}$ between 200 and 
400~km and with the average filling fraction set to 0.7.

The resultant log of the pdf of the simulation is given as the background
colour scale on Figure~\ref{ips}. The fit is acceptable, the only two
outliers are the interpulse width for PSR~J1057--5226 and PSR~J1825--0925.
For PSR~J1057--5226, $\alpha$ is as low as $70\degr$ and $s$ may be larger
than 1 \citep{ww09}.
PSR~J1825--0925 which may be aligned and not an orthogonal rotator
(see the arguments and counterarguments in \citealt{dzg05} and
\citealt{hkh+17}).
The simulation is also able to reproduce the difference in widths between
the main and interpulses; this is largely due to the differing
values of $\beta$ between the two lines-of-sight, something which
\citet{mgr11} do not take into account.

\subsection{Extension to the whole dataset}
Given the result for the interpulse pulsars, we fix the emission
height range and the filling factor to those used in Figure~\ref{ips}.
To extend these results to the entire dataset,
we need an $\alpha$ distribution and we therefore
assume that the {\it intrinsic} distribution is random.  Because of
the effects of beaming, the {\it observed} distribution is sinusoidal
(i.e. it peaks at $\alpha=90\degr$).
The results are shown in Figure~\ref{norook}.

It is immediately noticeable that although the bulk of the pulsars
fall within the highest contours of the pdf, the observed distribution
has a higher mean than the simulation and there are a substantial
fraction of pulsars which are significantly wider observationally than expected
in the simulation. Furthermore, the simulation predicts a much larger
number of interpulses than seen in the observational data.
These deviations mean that the assumptions going into producing
the simulation are not correct; either the $\alpha$ distribution is not
random, and/or there is some dependence with $s$ on $\alpha$.
\begin{figure}
\includegraphics[width=6.5cm,angle=-90]{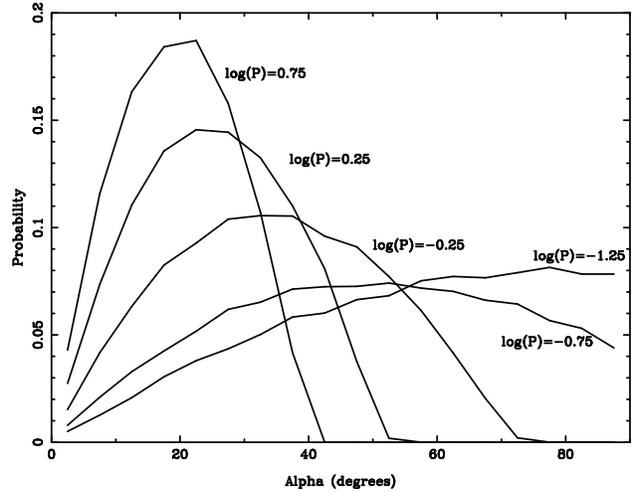}
\caption{Probability density function of $\alpha$ shown for 5 values of log($P$)
in equal steps from $-1.25$ to $+0.75$.}
\label{alpha}
\end{figure}

\subsection{Decay of $\alpha$ with time}
If $\alpha$ decays with time (as Figures~\ref{edot} and \ref{norook}
seem to hint at) then older pulsars 
should also have larger pulse widths as a result, relative to the no decay case.
Problematically we cannot determine the true age of a pulsar as the so-called
characteristic age, $\tau_c = P/(2\dot{P})$, has been shown not to be a good indicator, especially for older pulsars as shown in JK17.
This makes it hard to include $\alpha$ decay in the simulations. To achieve this, we must find a way to use $P$ as
a proxy for age and therefore to establish $\alpha$ as a function of $P$. We can draw on the results of the simulation in JK17,
who incorporated $\alpha$-decay and were able to match their simulated
pulsar population with the observed population. This
gives us the probability density function of $\alpha$ as a function of $P$ for the simulated pulsars that also pass the detection criteria.

\begin{figure}
\includegraphics[width=6.5cm,angle=-90]{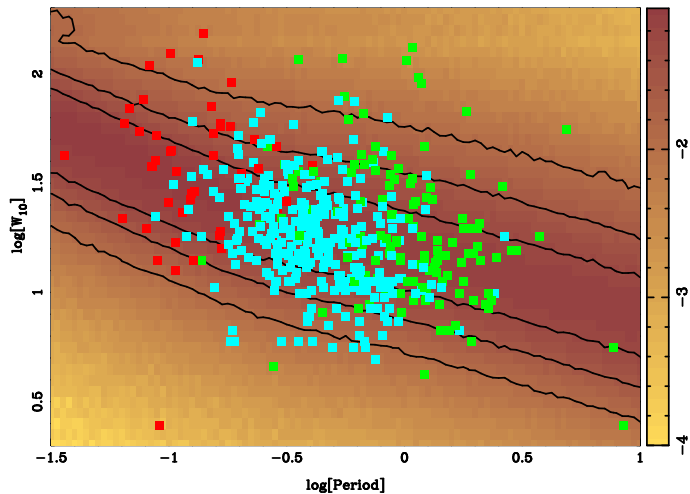}
\caption{As for Figure~\ref{norook} with the addition of an $\alpha$
distribution which depends on $P$ (see text and Figure~\ref{alpha} for
details).  Squares denote the measurements with the colours representing 
$\dot{E}$ as in Figure~\ref{data-all}.
Colour coding of the background and the contours are as per Figure~\ref{ips}.}
\label{all}
\end{figure}

To assess what happens if we include $\alpha$ decay in this way into the current simulation,
we convolve the original random distribution in $\alpha$ with a cut-off
which depends on $P^2$.  The resultant probability density function of $\alpha$
for 5 different spin periods is shown in Figure~\ref{alpha}.
These distributions are consistent with those found for pulsars with different
pulse periods in the work of JK17.
Including this form of $\alpha$-decay in the simulations here leads to the 
results shown in Figure~\ref{all}.
The main difference between Figures~\ref{all} and \ref{norook}
is that the pdf bends upwards at large $P$ as the $\alpha$ distribution
is skewed towards smaller values at these periods.

\begin{figure}
\includegraphics[width=6.5cm,angle=-90]{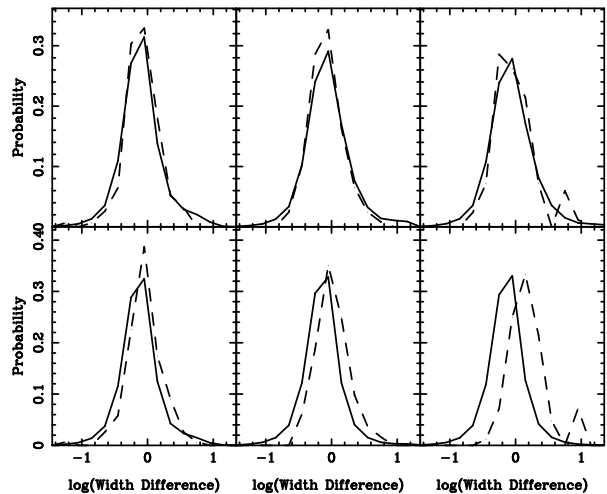}
\caption{Histogram of the difference between the simulated (or observed)
pulse width versus the expected pulse width for different values of $P$.
Bottom panels: Results from the case without $\alpha$-decay
see Figure~\ref{norook}. Top panels: Results from the case with
$\alpha$-decay; see Figure~\ref{all}. From left to right, panels represent
pulsars with $P<0.3$~s, $0.3<P<1.0$~s, $P>1.0$~s, the solid line represents the 
simulation and the dashed line the observed population.}
\label{compare}
\end{figure}

We assess the requirement to include $\alpha$ decay in the following way. First we compute
the peak of the simulated pulse-width distribution for each $P$ in the range we have simulated, i.e. the most likely pulse width for each period.
We subtract the pulse widths for both the simulation and
the observed distribution from this peak value. This has the effect of
removing the correlation between $W$ and $P$ allowing us to compare
the distribution across period. We bin the data into 3 bins across the period
axis, $P<0.3$~s, $0.3<P<1.0$~s, $P>1.0$~s.
Figure~\ref{compare} shows the histograms of the resulting distributions
for the simulation and the observations for the cases without and with
$\alpha$-decay. Clearly, without $\alpha$-decay, the match between the
histograms is poor especially for pulsars with long periods,
whereas with $\alpha$-decay the simulation and data match well in all 3 period bins.

\subsection{Dependence of $s$ on $\alpha$}
It is already well established that high $\dot{E}$ pulsars have
wide profiles \citep{jw06} and these could be caused by high emission
heights \citep{kj07,wj08b} or emission from outside the conventional
polar cap region \citep{rwj15b}.
\citet{rwj15b} have derived a dependence for $s$ on $\alpha$ 
via the equation
\begin{equation}
s = -0.022 \alpha + 2.80
\label{s-alpha}
\end{equation}
with $\alpha$ in degrees. We note that this maintains $s\sim 1$
when $\alpha\sim 90\degr$ and so this does not change the results
for the interpulses given in Figure~\ref{ips}. However, for low
values of $\alpha$ it allows $s$ to be significantly larger than
1 which results in a larger distribution of possible widths.
At the same time, \citet{rwj15b} recognised that the filling factor of
the beams was small and that many components were ``missing" which meant
that the `true' widths they assumed were significantly larger than the
observed widths. It is not clear how the results from their sample of 
high $\dot{E}$ gamma-ray pulsars can be extrapolated to 
the long-period, low $\dot{E}$ bulk of the pulsar population. We do not
include this effect in our simulations.

\subsection{Singles and multiples}
What are the implications of the relative fraction of singles and multiples
and the differences between the widths of single and multipe component pulsars?

In a simple version of the Rankin beam models, the numbers can be reconciled
if the core component is relatively large and the the conal component occupies 
a large fraction of the beam.  In this case, a large fraction
of single components can be obtained by cutting through the cone at
moderate to high $\beta$ and these single components will also be relatively 
wide compared to double components at lower $\beta$.
In addition, narrow single components can be obtained where the cone is not 
present and the line of sight cuts through the core component.

In the patchy model of \citet{lm88}, the number of active patches along
the line of sight determines the profile shape. In order to have an 
equal fraction of single and multiple profile pulsars, the patch size must
be relatively large and should overlap frequently. This seems consistent
with the mean beam emissivity shown by \citet{hm01}.

\section{Summary}
We have used the database of pulsars provided by \citet{jk18} and a novel
method based on a Gaussian Process technique to determine the pulse
widths at the 10\% level for a sample of 600 pulsars at an observing
frequency of 1.4~GHz. We have investigated the relationship between a
pulsar's spin period and its width and find that:
\begin{itemize}
\item for a given spin period, a broad range of widths are observed, although
the general trend is that $W_{10}\propto P^{-0.3}$.
\item emission heights are low and have a narrow range between 200
and 400~km irrespective of $P$.
\item pulsars which have interpulse emission and are orthogonal rotators show
evidence for an underfilled beam; i.e. their pulse widths are narrower than
expected.
\item pulsars with low $\dot{E}$ have wider profiles than expected from
their spin periods; this is most likely due to $\alpha$ decay in this older
population.
\item there should exist a population of long-period, low-$\alpha$ pulsars
which are difficult to detect with standard techniques but which may be
amenable to discover via a fast-folding algorithm \citep{cbc+17}.
\item pulsars with multiple components are wider than those with only
one component. This has implications for the overall beam structure
and the size of the emitting regions, but it remains difficult to
distinguish between the models of \citet{ran93} and \citet{lm88}.
\end{itemize}

\section*{Acknowledgments}
We used the ATNF pulsar catalogue at 
\url{http://www.atnf.csiro.au/people/pulsar/psrcat/}
for this work.
The Parkes telescope is part of the Australia Telescope National Facility
which is funded by the Commonwealth of Australia for operation as a
National Facility managed by CSIRO.

\bibliographystyle{mnras}
\bibliography{pw}

\begin{thebibliography}{}

\bibitem[\protect\citeauthoryear{{Blaskiewicz}, {Cordes}, \&
  {Wasserman}}{{Blaskiewicz} et~al.}{1991}]{bcw91}
{Blaskiewicz} M., {Cordes} J.~M.,  {Wasserman} I., 1991, ApJ, 370, 643

\bibitem[\protect\citeauthoryear{{Cameron} et~al.}{{Cameron}
  et~al.}{2017}]{cbc+17}
{Cameron} A.~D., {Barr} E.~D., {Champion} D.~J., {Kramer} M.,  {Zhu} W.~W.,
  2017, MNRAS, 468, 1994

\bibitem[\protect\citeauthoryear{{Clifton} \& {Weisberg}}{{Clifton} \&
  {Weisberg}}{2008}]{cw08}
{Clifton} T.,  {Weisberg} J.~M., 2008, ApJ, 679, 687

\bibitem[\protect\citeauthoryear{{Desvignes} et~al.}{{Desvignes}
  et~al.}{2013}]{dkc+13}
{Desvignes} G., {Kramer} M., {Cognard} I., {Kasian} L., {van Leeuwen} J.,
  {Stairs} I.,  {Theureau} G., 2013, in IAU Symposium, Vol. 291, {van Leeuwen}
  J., ed, Neutron Stars and Pulsars: Challenges and Opportunities after 80
  years, p. 199

\bibitem[\protect\citeauthoryear{{Dyks}, {Zhang}, \& {Gil}}{{Dyks}
  et~al.}{2005}]{dzg05}
{Dyks} J., {Zhang} B.,  {Gil} J., 2005, ApJ, 626, L45

\bibitem[\protect\citeauthoryear{{Fonseca}, {Stairs}, \& {Thorsett}}{{Fonseca}
  et~al.}{2014}]{fst14}
{Fonseca} E., {Stairs} I.~H.,  {Thorsett} S.~E., 2014, ApJ, 787, 82

\bibitem[\protect\citeauthoryear{{Gangadhara}}{{Gangadhara}}{2004}]{gan04}
{Gangadhara} R.~T., 2004, ApJ, 609, 335

\bibitem[\protect\citeauthoryear{{Gil}, {Gronkowski}, \& {Rudnicki}}{{Gil}
  et~al.}{1984}]{ggr84}
{Gil} J., {Gronkowski} P.,  {Rudnicki} W., 1984, A\&A, 132, 312

\bibitem[\protect\citeauthoryear{{Gil}, {Kijak}, \& {Seiradakis}}{{Gil}
  et~al.}{1993}]{gks93}
{Gil} J.~A., {Kijak} J.,  {Seiradakis} J.~H., 1993, A\&A, 272, 268

\bibitem[\protect\citeauthoryear{{Gould} \& {Lyne}}{{Gould} \&
  {Lyne}}{1998}]{gl98}
{Gould} D.~M.,  {Lyne} A.~G., 1998, MNRAS, 301, 235

\bibitem[\protect\citeauthoryear{{Gupta} \& {Gangadhara}}{{Gupta} \&
  {Gangadhara}}{2003}]{gg03}
{Gupta} Y.,  {Gangadhara} R.~T., 2003, ApJ, 584, 418

\bibitem[\protect\citeauthoryear{{Han} \& {Manchester}}{{Han} \&
  {Manchester}}{2001}]{hm01}
{Han} J.~L.,  {Manchester} R.~N., 2001, MNRAS, 320, L35

\bibitem[\protect\citeauthoryear{{Hermsen} et~al.}{{Hermsen}
  et~al.}{2017}]{hkh+17}
{Hermsen} W. et~al., 2017, MNRAS, 466, 1688

\bibitem[\protect\citeauthoryear{{Johnston} \& {Karastergiou}}{{Johnston} \&
  {Karastergiou}}{2017}]{jk17}
{Johnston} S.,  {Karastergiou} A., 2017, MNRAS, 467, 3493

\bibitem[\protect\citeauthoryear{{Johnston} \& {Kerr}}{{Johnston} \&
  {Kerr}}{2018}]{jk18}
{Johnston} S.,  {Kerr} M., 2018, MNRAS, 474, 4629

\bibitem[\protect\citeauthoryear{{Johnston} \& {Weisberg}}{{Johnston} \&
  {Weisberg}}{2006}]{jw06}
{Johnston} S.,  {Weisberg} J.~M., 2006, MNRAS, 368, 1856

\bibitem[\protect\citeauthoryear{{Karastergiou} \& {Johnston}}{{Karastergiou}
  \& {Johnston}}{2007}]{kj07}
{Karastergiou} A.,  {Johnston} S., 2007, MNRAS, 380, 1678

\bibitem[\protect\citeauthoryear{{Keith} et~al.}{{Keith} et~al.}{2008}]{kjk+08}
{Keith} M.~J., {Johnston} S., {Kramer} M., {Weltevrede} P., {Watters} K.~P.,
  {Stappers} B.~W., 2008, MNRAS, 389, 1881

\bibitem[\protect\citeauthoryear{{Keith} et~al.}{{Keith} et~al.}{2010}]{kjwk10}
{Keith} M.~J., {Johnston} S., {Weltevrede} P.,  {Kramer} M., 2010, MNRAS, 402,
  745

\bibitem[\protect\citeauthoryear{{Kramer} \& {Johnston}}{{Kramer} \&
  {Johnston}}{2008}]{kj08}
{Kramer} M.,  {Johnston} S., 2008, MNRAS, 390, 87

\bibitem[\protect\citeauthoryear{{Kramer} et~al.}{{Kramer}
  et~al.}{1994}]{kwj+94}
{Kramer} M., {Wielebinski} R., {Jessner} A., {Gil} J.~A.,  {Seiradakis} J.~H.,
  1994, A\&AS, 107

\bibitem[\protect\citeauthoryear{{Kramer} et~al.}{{Kramer}
  et~al.}{1998}]{kxl+98}
{Kramer} M., {Xilouris} K.~M., {Lorimer} D.~R., {Doroshenko} O., {Jessner} A.,
  {Wielebinski} R., {Wolszczan} A.,  {Camilo} F., 1998, ApJ, 501, 270

\bibitem[\protect\citeauthoryear{{Lazarus} et~al.}{{Lazarus}
  et~al.}{2015}]{lbh+15}
{Lazarus} P. et~al., 2015, ApJ, 812, 81

\bibitem[\protect\citeauthoryear{{Lyne} et~al.}{{Lyne} et~al.}{2013}]{lgw+13}
{Lyne} A., {Graham-Smith} F., {Weltevrede} P., {Jordan} C., {Stappers} B.,
  {Bassa} C.,  {Kramer} M., 2013, Science, 342, 598

\bibitem[\protect\citeauthoryear{{Lyne} \& {Manchester}}{{Lyne} \&
  {Manchester}}{1988}]{lm88}
{Lyne} A.~G.,  {Manchester} R.~N., 1988, MNRAS, 234, 477

\bibitem[\protect\citeauthoryear{{Maciesiak} \& {Gil}}{{Maciesiak} \&
  {Gil}}{2011}]{mg11}
{Maciesiak} K.,  {Gil} J., 2011, MNRAS, 417, 1444

\bibitem[\protect\citeauthoryear{{Maciesiak}, {Gil}, \&
  {Melikidze}}{{Maciesiak} et~al.}{2012}]{mgm12}
{Maciesiak} K., {Gil} J.,  {Melikidze} G., 2012, MNRAS, 424, 1762

\bibitem[\protect\citeauthoryear{{Maciesiak}, {Gil}, \& {Ribeiro}}{{Maciesiak}
  et~al.}{2011}]{mgr11}
{Maciesiak} K., {Gil} J.,  {Ribeiro} V.~A.~R.~M., 2011, MNRAS, 414, 1314

\bibitem[\protect\citeauthoryear{{Manchester} et~al.}{{Manchester}
  et~al.}{2010}]{mks+10}
{Manchester} R.~N. et~al., 2010, ApJ, 710, 1694

\bibitem[\protect\citeauthoryear{{McKinnon}}{{McKinnon}}{1993}]{mck93}
{McKinnon} M.~M., 1993, ApJ, 413, 317

\bibitem[\protect\citeauthoryear{{Mitra} \& {Rankin}}{{Mitra} \&
  {Rankin}}{2002}]{mr02}
{Mitra} D.,  {Rankin} J.~M., 2002, ApJ, 577, 322

\bibitem[\protect\citeauthoryear{{Mitra} \& {Rankin}}{{Mitra} \&
  {Rankin}}{2011}]{mr11}
{Mitra} D.,  {Rankin} J.~M., 2011, ApJ, 727, 92

\bibitem[\protect\citeauthoryear{{Narayan} \& {Vivekanand}}{{Narayan} \&
  {Vivekanand}}{1983}]{nv83}
{Narayan} R.,  {Vivekanand} M., 1983, A\&A, 122, 45

\bibitem[\protect\citeauthoryear{{Rankin}}{{Rankin}}{1990}]{ran90}
{Rankin} J.~M., 1990, ApJ, 352, 247

\bibitem[\protect\citeauthoryear{{Rankin}}{{Rankin}}{1993}]{ran93}
{Rankin} J.~M., 1993, ApJ, 405, 285

\bibitem[\protect\citeauthoryear{{Ravi}, {Manchester}, \& {Hobbs}}{{Ravi}
  et~al.}{2010}]{rmh10}
{Ravi} V., {Manchester} R.~N.,  {Hobbs} G., 2010, ApJ, 716, L85

\bibitem[\protect\citeauthoryear{{Roberts} et~al.}{{Roberts}
  et~al.}{2012}]{roe+12}
{Roberts} S., {Osborne} M., {Ebden} M., {Reece} S., {Gibson} N.,  {Aigrain} S.,
  2012, Philosophical Transactions of the Royal Society of London Series A,
  371, 20110550

\bibitem[\protect\citeauthoryear{{Rookyard}, {Weltevrede}, \&
  {Johnston}}{{Rookyard} et~al.}{2015a}]{rwj15a}
{Rookyard} S.~C., {Weltevrede} P.,  {Johnston} S., 2015a, MNRAS, 446, 3367

\bibitem[\protect\citeauthoryear{{Rookyard}, {Weltevrede}, \&
  {Johnston}}{{Rookyard} et~al.}{2015b}]{rwj15b}
{Rookyard} S.~C., {Weltevrede} P.,  {Johnston} S., 2015b, MNRAS, 446, 3356

\bibitem[\protect\citeauthoryear{{Ruderman} \& {Sutherland}}{{Ruderman} \&
  {Sutherland}}{1975}]{rs75}
{Ruderman} M.~A.,  {Sutherland} P.~G., 1975, ApJ, 196, 51

\bibitem[\protect\citeauthoryear{{Skrzypczak} et~al.}{{Skrzypczak}
  et~al.}{2018}]{sbm+18}
{Skrzypczak} A., {Basu} R., {Mitra} D., {Melikidze} G.~I., {Maciesiak} K.,
  {Koralewska} O.,  {Filothodoros} A., 2018, ApJ, 854, 162

\bibitem[\protect\citeauthoryear{{Szary} \& {van Leeuwen}}{{Szary} \& {van
  Leeuwen}}{2017}]{svl17}
{Szary} A.,  {van Leeuwen} J., 2017, ApJ, 845, 95

\bibitem[\protect\citeauthoryear{{Tauris} \& {Manchester}}{{Tauris} \&
  {Manchester}}{1998}]{tm98}
{Tauris} T.~M.,  {Manchester} R.~N., 1998, MNRAS, 298, 625

\bibitem[\protect\citeauthoryear{{van Heerden}, {Karastergiou}, \&
  {Roberts}}{{van Heerden} et~al.}{2017}]{vhkr17}
{van Heerden} E., {Karastergiou} A.,  {Roberts} S.~J., 2017, MNRAS, 467, 1661

\bibitem[\protect\citeauthoryear{{Wang} et~al.}{{Wang} et~al.}{2014}]{wpz+14}
{Wang} H.~G. et~al., 2014, ApJ, 789, 73

\bibitem[\protect\citeauthoryear{{Weltevrede} \& {Johnston}}{{Weltevrede} \&
  {Johnston}}{2008a}]{wj08b}
{Weltevrede} P.,  {Johnston} S., 2008a, MNRAS, 391, 1210

\bibitem[\protect\citeauthoryear{{Weltevrede} \& {Johnston}}{{Weltevrede} \&
  {Johnston}}{2008b}]{wj08a}
{Weltevrede} P.,  {Johnston} S., 2008b, MNRAS, 387, 1755

\bibitem[\protect\citeauthoryear{{Weltevrede} \& {Wright}}{{Weltevrede} \&
  {Wright}}{2009}]{ww09}
{Weltevrede} P.,  {Wright} G., 2009, MNRAS, 395, 2117

\bibitem[\protect\citeauthoryear{{Wright} \& {Weltevrede}}{{Wright} \&
  {Weltevrede}}{2017}]{ww17}
{Wright} G.,  {Weltevrede} P., 2017, MNRAS, 464, 2597

\bibitem[\protect\citeauthoryear{{Young} et~al.}{{Young} et~al.}{2010}]{ycbb10}
{Young} M.~D.~T., {Chan} L.~S., {Burman} R.~R.,  {Blair} D.~G., 2010, MNRAS,
  402, 1317

\end{thebibliography}
\bsp
\label{lastpage}
\end{document}